\documentclass{rnaastex}

\usepackage{amsmath}
\usepackage{url}
\usepackage{xspace}
\usepackage{placeins}

\newcommand{\sdssr}{SDSS \textsl{r}\xspace}
\newcommand{\kepler}{\textsl{Kepler}\xspace}

\begin{document}

\title{Large Starspot Groups on HAT-P-11 in Activity Cycle 1}

\author[0000-0003-2528-3409]{Brett M. Morris}
\affiliation{Astronomy Department, University of Washington, Seattle, WA 98195, USA}

\author{Suzanne L. Hawley}
\affiliation{Astronomy Department, University of Washington, Seattle, WA 98195, USA}

\author{Leslie Hebb}
\affiliation{Physics Department, Hobart and William Smith Colleges, Geneva, NY 14456, USA}

\email{bmmorris@uw.edu}

\vspace{-2cm}

\FloatBarrier\section{Background}
HAT-P-11 is a planet-hosting K4V star in the \kepler field, with an activity cycle that bear similarities to the Sun's \citep{Bakos2010, sanchis-ojeda2011, Morris2017a}. The chromospheric activity of HAT-P-11 passed through minimum in 2016, and seems to be rising back towards maximum \citep{Morris2017b}. In late 2017, HAT-P-11's $S$-index had returned to $S\sim0.55$, similar to the activity level in the previous cycle (Cycle 0). We report ground-based observations to measure the starspots of HAT-P-11 in its second observed magnetic activity cycle (hereafter Cycle 1). Spot occultations have not yet been observed at the present phase of HAT-P-11's activity cycle. 

\section{Observations}

We gathered photometry of HAT-P-11 and six comparison stars through a holographic diffuser on the ARCTIC imager mounted on the Astrophysical Research Consortium (ARC) 3.5 m Telescope at Apache Point Observatory (APO) \citep{Huehnerhoff2016}. The diffuser enables precision photometry from the ground \citep{Stefansson2017}. We observed a transit of HAT-P-11 b using 2$\times$2 binning 10 second exposures in \sdssr on 2017 October 30 UTC. We normalize the HAT-P-11 light curve by a mean comparison star, which is computed from a linear combination of the following regressors: the fluxes of each comparison star, the target centroid pixel $x$ and $y$ coordinates, median sky background, air humidity, air pressure, and airmass -- see \citet{Morris2018a} for detailed explanation of the photometric technique. The standard deviation of the target flux before transit is 400 ppm in one minute bins, $\sim4\times$ larger than \kepler's photometric precision (90 ppm).

\section{Starspot properties}

\begin{figure}
\begin{center}
\begin{minipage}{6in}
  \centering
  \raisebox{-0.5\height}{\includegraphics[scale=0.6]{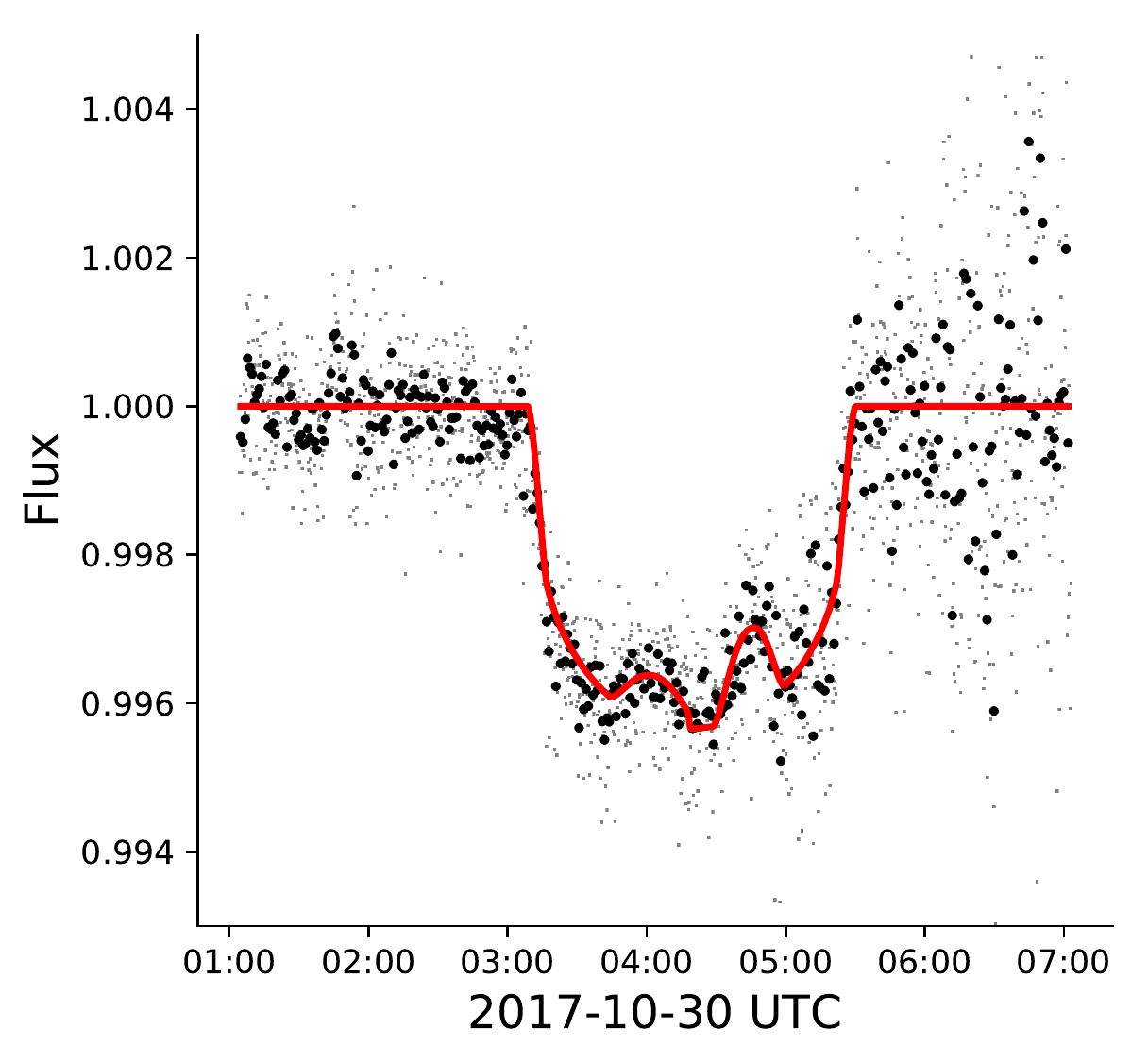}}
  \hspace*{.2 in}
  \raisebox{-0.5\height}{\includegraphics[scale=0.35]{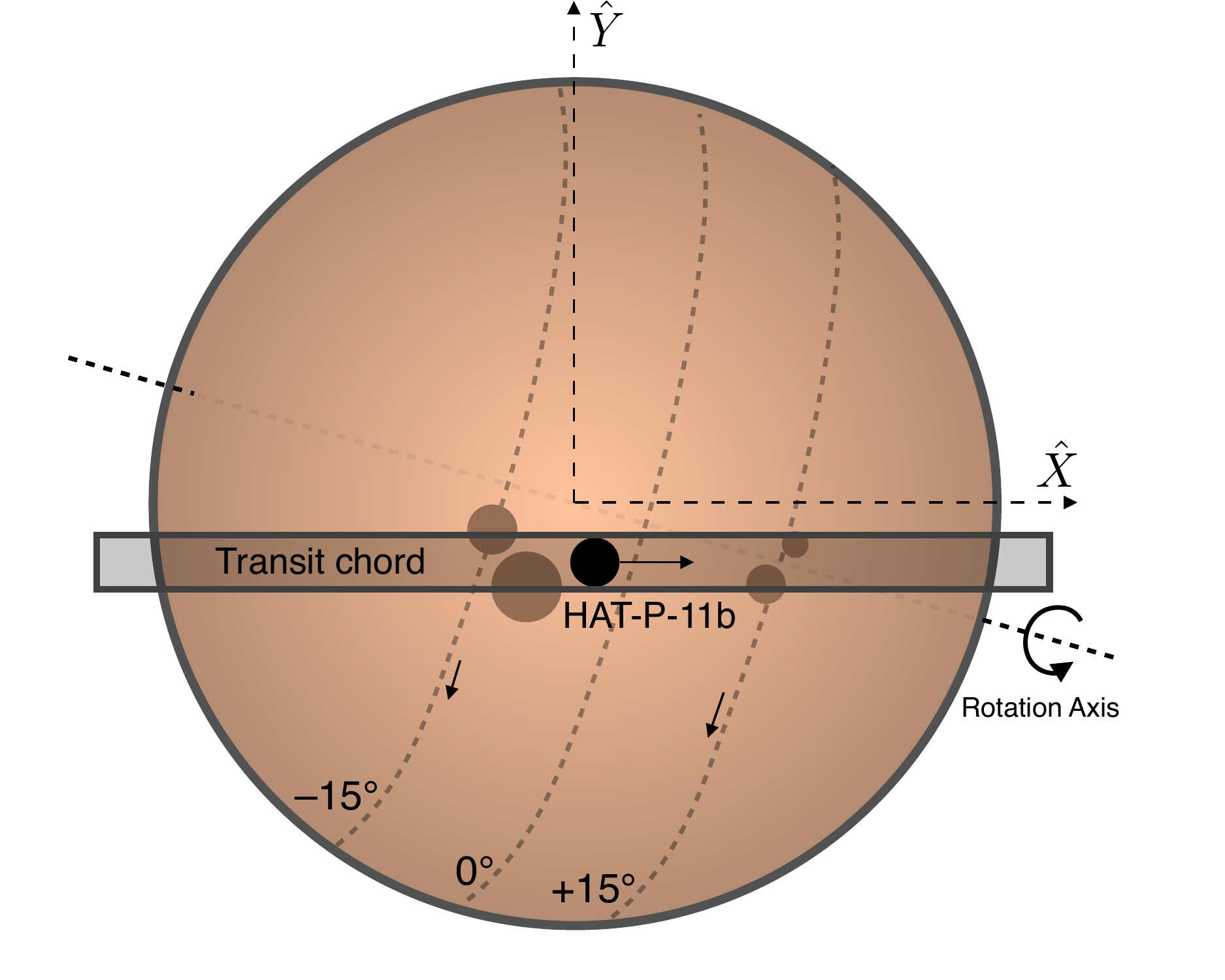}}
\end{minipage}
\caption{\textsl{Left}: Transit light curve at 10 second cadence (gray squares) and one minute bins (black circles), with the maximum-likelihood spot model from \texttt{STSP} (red curve). The scatter increases at later times due to high airmass. \textsl{Right}: schematic illustration of a possible spot group configuration consistent with the APO occultation photometry. See Figure 3 of \citet{Morris2017a} for a more detailed explanation of the notation.
\label{fig}}
\end{center}
\end{figure}

In Figure~\ref{fig}, the transit light curve shows a occultation of a large dark feature in the southern hemisphere before mid-transit, and a smaller dark feature in the northern hemisphere after mid-transit. We noted in \citet{Morris2017a} that adjacent spot occultations could give the appearance of much larger spots, and we emphasize here that each of these apparent spot occultations could in fact be occultations of groups of spots.  If these anomalies are in fact occultations of individual spots, they would be among the largest observed on HAT-P-11. However, given the similarities between the spots observed during the \kepler years \citep{Morris2017a} and sunspots \citep{Solanki2003}, we suggest that we are likely observing occultations of large spot groups appearing at the beginning of Cycle 1. Spot occultations have not previously been observed at this phase of the HAT-P-11 activity cycle.  

Fits with the \texttt{STSP} photometric spot model \citep{Morris2017a, Hebb2018} reveal the area coverage of spots within the transit chord for UTC 2017-10-30 is 14\% --- which makes this transit the most spotted HAT-P-11 transit observed to date. From 2009-2013, we measured the fraction of the transit chord occupied by spots in \citet{Morris2017a} to vary between 0.5-10\%. 

\acknowledgments

We thank Gudmundur Stefansson and Yiting Li for productive conversations. Based on observations obtained with the APO 3.5-meter telescope, which is owned and operated by ARC.

\software{\texttt{astropy} \citep{Astropy2013}, \texttt{photutils} \citep{Bradley2016}, \texttt{friedrich} \citep{Morris2017a}, \texttt{astroplan} \citep{astroplan}, STSP \citep{Hebb2018}, \texttt{astroscrappy} \citep{astroscrappy}}

\facility{APO/ARC 3.5m}


\end{document}